\documentclass[journal]{IEEEtran}

\ifCLASSINFOpdf
\else
\fi
\makeatletter
\newcommand{\rmnum}[1]{\romannumeral #1}
\newcommand{\Rmnum}[1]{\expandafter\@slowromancap\romannumeral #1@}
\makeatother
\usepackage{graphics}
\usepackage{graphicx}
\usepackage{subfigure}
\usepackage{epsfig}
\usepackage{amsmath}
\usepackage{amssymb}
\usepackage{amsfonts}
\usepackage{amsthm}
\usepackage[ruled,vlined,linesnumbered]{algorithm2e}
\usepackage{algorithmic}
\usepackage{multirow}
\usepackage{multicol}
\usepackage{bm}
\usepackage{cite}
\usepackage{balance}
\usepackage{color}
\usepackage{booktabs}
\usepackage{longtable}
\usepackage{epstopdf}
\usepackage{stfloats}
\usepackage{hyperref}
\usepackage{threeparttable}
\usepackage{makecell}

\usepackage[mathscr]{eucal}
\hypersetup{hidelinks,
	colorlinks=true,
	allcolors=black,
	pdfstartview=Fit,
	breaklinks=true}
\begin{document}
\title{Meta-learning based Selective Fixed-filter Active Noise Control System with ResNet Classifier}
\author{{Yingying~Xiao,
       Meiqin~Liu,~\IEEEmembership{Senior Member,~IEEE},
       Wei Dai,~\IEEEmembership{Senior Member, IEEE}, and Jian Lan, \IEEEmembership{Senior Member, IEEE}
\vspace{-10pt}}
\thanks{This work was supported by the National Natural Science Foundation of China under Grant U23B2035 and the Fundamental Research Funds for Xi'an Jiaotong University under Grant xtr072022001. \emph{(Corresponding author: Meiqin Liu.)}}
\thanks{Yingying Xiao and Jian Lan are with the National Key Laboratory of Human-Machine Hybrid Augmented Intelligence, Xi'an
 Jiaotong University, Xi'an 710049, China (email: xyy2024@stu.xjtu.edu.cn and lanjian@xjtu.edu.cn).}
\thanks{Meiqin Liu is with the National Key Laboratory of Human-Machine Hybrid Augmented Intelligence, Xi'an Jiaotong University, Xi'an 710049, China and also with the College of Electrical Engineering, Zhejiang University, Hangzhou 310027, China (e-mail:liumeiqin@zju.edu.cn).}
\thanks{Wei Dai is with the School of Information and Control Engineering, China University of Mining and Technology, Xuzhou 221116, China (e-mail: weidai@cumt.edu.cn).}
}
\maketitle

\begin{abstract}
The selective fixed-filter strategy is popular in industrial applications involving active noise control (ANC) technology, which circumvents the time-consuming online learning process by selecting the best-matched pre-trained control filter. However, the existing selective fixed-filter ANC (SFANC) based algorithms classify noises in frequency band, which is not a reasonable approach. Moreover, they pre-train the control filter utilizing only a single noise segment, leading to inaccurate estimation and undesirable noise cancellation performance when dealing with dynamically time-varying noise. Inspired by the applicability of meta-learning to various models utilizing gradient descent technique, this paper proposes a novel meta-learning based SFANC system, wherein the fixed-filters that may not be optimal for specific types of noises but can rapidly adapt to previously unseen noise conditions are pre-trained. To address the mismatch issue between meta-learning update methods and ANC requirements while enhancing the receptive field and convergence speed of control filters, a multiple-input batch processing strategy is utilized in pre-training. Simulations based on the common ESC-50 noise dataset are performed and demonstrate the superiorities of the proposed method in terms of classification accuracy, convergence speed, and steady-state noise cancellation.
\end{abstract}

\begin{IEEEkeywords}
Selective fixed-filter active noise control, meta-learning, time-varying noise, residual net.
\end{IEEEkeywords}
\maketitle

\section{Introduction}
\IEEEPARstart{A}ctive noise control (ANC) technology can effectively eliminate the disturbance noise by generating an anti-noise that is equal in amplitude and opposite in phase to the disturbance \cite{FxLMS,ANC1,Survey,FxNLMS}, and has been widely used in headphones \cite{headphone1,headphone2}, intelligent driving \cite{Vehical}, traffic noise control \cite{traffic}, etc. Numerous noise control algorithms have been developed over the past few decades, including linear noise control algorithms such as the well-known filtered-x least mean square (FxLMS) \cite{FxLMS} and normalized FxLMS (FxNLMS) algorithms \cite{FxNLMS}, as well as nonlinear noise control algorithms \cite{VFxLMS,NonlinearP,CheyFLNN,NysMANC,FsLMS,AEFLNN,KFxLMS,RFFxLMS,RFCGFxGHT} like the Volterra filtered-x least mean square (VFxLMS) \cite{VFxLMS} and filtered-s least mean square (FsLMS) algorithms \cite{FsLMS}. However, these online learning based adaptive control algorithms require a long response time for time-varying noise, which is unacceptable in practical applications, e.g., ANC headphones \cite{headphone1}.

Recently, some studies in this domain have looked beyond traditional online learning methods to selective fixed-filter active noise control (SFANC) methods to shorten the filter response time \cite{headphone2,SelectiveFrequency,SelectiveCNN,SelectiveTransfer,SFANC-FxNLMS}. The lately presented hybrid SFANC-FxNLMS algorithm \cite{SFANC-FxNLMS} can provide fast convergence speed and continuous updating ability by combining SFANC \cite{SelectiveCNN} and FxNLMS \cite{FxNLMS}. However, since real-life noises typically span a wide frequency range, it may be improper to classify noises according to frequency range in these SFANC based methods. Moreover, these SFANC based methods pre-train the control filter utilizing only a single noise segment with the FxNLMS algorithm, failing to learn the statistical characteristics of all noises belonging to the same category. That finally leads to inaccurate control filter estimation and large steady-state errors when dealing with time-varying noise.

As a model-agnostic method applicable to any model trained with the gradient descent rule, meta-learning \cite{meta1,meta2} not only drives the advancement of deep learning but also provides new insights to ANC technology. The modified model-agnostic meta-learning (MAML) methods for initialization are proposed for both single channel \cite{MAML-ANC} and multichannel ANC scenarios \cite{MAML-MANC} to accelerate the algorithm's initial convergence speed. Unlike the existing MAML initialization methods, which focus on the problem of slow initial convergence caused by initial zero-padded input vectors passing through finite impulse response filters, we focus more on how meta-learning can be used to train control filters that can quickly learn and adapt to a human-commonsense homogeneous with training noise but unseen time-varying noise.

\begin{figure}[!t]
\centerline{\includegraphics[width=9cm,height=3.3cm]{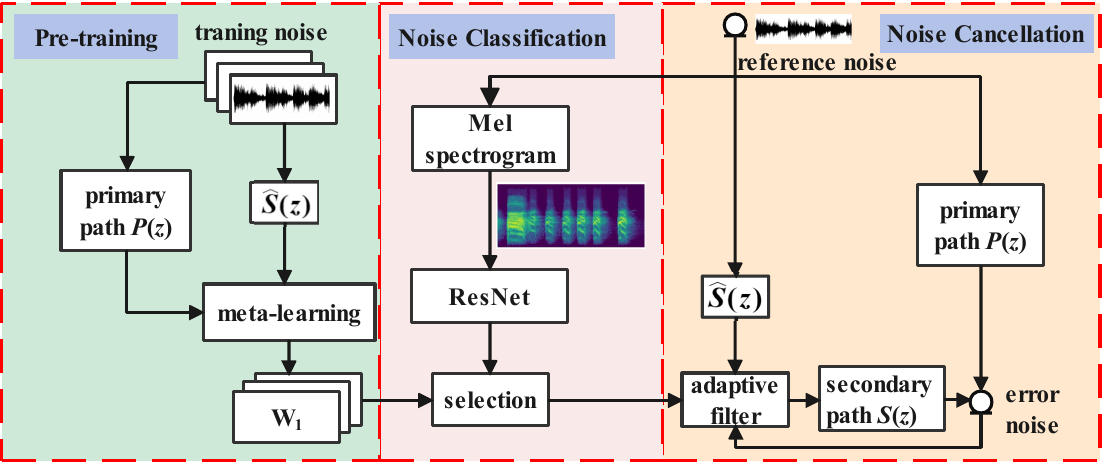}}
\caption{Diagram of the proposed meta-learning based SFANC system.}
\label{Figure1} 
\vspace{-10pt}
\end{figure}
\raggedbottom
Specifically, this paper proposes a novel ANC framework including the pre-training, noise classification, and noise cancellation modules, as Fig. \ref{Figure1} shows. In the pre-training module, the control filter database trained by the proposed MAML-FxLMS algorithm is built. Then, the arriving reference noise is classified into a noise category by a residual net (ResNet) classifier, and the corresponding control filter is selected and updated in the next noise cancellation module. Benefited from the excellence for learning to learn of meta-learning, each pre-trained control filter can quickly approach the optimal filter within few iterations for various noise environments, that is what other SFANC based algorithms lack.

\section{The meta-learning based pre-training method}
The noise control process of the well-known FxLMS \cite{SFANC-FxNLMS} algorithm can be described as
\begin{equation}
\left\{ {\begin{array}{*{20}{l}}
{e(n) = d(n) - [{{\bf{w}}^{\rm{T}}}(n){\bf{x}}(n)]*s(n)}\\
{{\bf{w}}(n) = {\bf{w}}(n) + \mu e(n){\bf{\tilde x}}(n)}
\end{array}} \right.
\end{equation}
where $n$ is the time index; $e(n)$ and $d(n)\in {{\mathbb{R}}}$ are the estimated error and the disturbance noise; ${\bf{x}}(n)$ and ${\bf{w}}(n) \in {{\mathbb{R}}}{^{L \times 1}}$ are the input vector constituted by reference noise $x(n)$ and the weight vector of the control filter with $L$ denoting the filter length, respectively; ${s}(n)$ is the impulsive response of the secondary path $S(z)$; ${\bf{\tilde x}}(n) = {\bf{x}}(n) * {\hat{s}}(n)$ with $*$ and ${\hat{s}}(n)$ being the linear convolution operation and the estimation of ${s}(n)$; $\mu$ is the step-size.

Consider a noise dataset consisting of $N$ noise categories, and each of them contains $M$ subclasses that are homologous in human-commonsense. For any noise category, we define the ANC task as $\mathcal{T}_{i} = \{\mathcal{L}{_{\mathcal{T}_i}(f(x_i))}, S(z), L \} $ for the $i$th subclass noise, where $\mathcal{L}{_{\mathcal{T}_i}(f(x_i))}$ is the loss with $f(x_i)$ representing a mapping from reference noise $x_i$ to the filter output $\hat{d}$; $S(z)$ and $L$ for all tasks are the same. We introduce a distribution over the tasks that belong to the same category as $p(\mathcal{T})$.

Rewrite $f(x_i)$ as $f_{\mathbf{w}_{i}}$ for clarity of expression of the parameterized model determined by parameter $\mathbf{w}_i$. Then, the update rule of $\mathbf{w}_i$ by adopting one or more gradient descent step on the generally used MSE loss \cite{FxLMS,FxNLMS} can be used to search the control filter. To provide the control filter with a larger receptive field and faster convergence speed in training, a multiple-input batch processing strategy is utilized by taking the set $\{{\bf{x}}_{i,k}^{\sup}(n),d_{i,k}^{\sup}(n) \}_{k=1}^K$ drawn from support set $D_i^{\sup}$ as input-output pairs for $\mathcal{T}_i$, and the one gradient descent step-based update rule is used for simplicity here. The updated control filter for task $\mathcal{T}_i$ at the iteration $n$ is denoted as ${\bf{w}}_i'(n)$, which is computed by
\begin{align}\label{eq2}
{\bf{w}}_i'(n)&= {\bf{w}}_i(n)-\alpha\nabla_{{\bf{w}}_i}\mathcal{L}_{\mathcal{T}_{i}}(f_{{\bf{w}}_i})\\
&={\bf{w}}_i(n) + \alpha {\bf{\tilde X}}_i^{\sup}(n){\left({\bf{e}}_i^{\sup}(n) \right)^{\rm{T}}}\nonumber
\end{align}
where ${\bf{\tilde X}}_i^{\sup}(n) = [{\bf{\tilde x}}_{i,1}^{\sup}(n),\cdots,{\bf{\tilde x}}_{i,K}^{\sup}(n)]$ with its $k$ column being ${\bf{\tilde x}}_{i,k}^{\sup}(n)={\bf{x}}_{i,k}^{\sup}(n)*\hat{s}(n)$; ${\bf{e}}_i^{\sup}(n) = {\bf{d}}_i^{\sup}(n) - [{{{\bf{w}}_i^{\rm{T}}}(n)}$ ${\bf{X}}_i^{\sup}(n)] *s(n)$ with ${\bf{d}}_i^{\sup}(n)=[d_{i,k}^{\sup}(n), \cdots, d_{i,K}^{\sup}(n)]$; $\alpha$ is the step size for inner-layer tasks.

Since the object model $f_{\mathbf{w}}$ with parameter $\mathbf{w}$ is expected to earn the ``learn to learn" ability by training across tasks following distribution $p(\mathcal{T})$ in meta-learning, the outer-layer $\mathbf{w}$ is transmitted into the inner-layer at each iteration, i.e., taking ${\bf{w}}_i(n)={\bf{w}}$ in (\ref{eq2}). And the loss of $f_{\mathbf{w}}$ based on the updated parameter ${\bf{w}}_i'(n)$ is given as
\begin{align}\label{eq3}
{\mathcal{L}}(f_{\mathbf{w}})=\sum_{\mathcal{T}_{i}\sim p(\mathcal{T})}\mathcal{L}_{\mathcal{T}_{i}}(f_{{\bf{w}}_i'(n)})
\end{align}
The set $\{ \{ {{\bf{x}}_{i,j}^{\rm{que}}(n),}$ ${d_{i,j}^{\rm{que}}(n)} \}_{j = 1}^J \}_{i = 1}^M $ drawn form query set $\{D_i^{\rm{que}}\}_{i = 1}^M $ is used as the input-output pairs to train the outer-layer filter $\mathbf{w}$. The MSE loss is adopted, allowing (\ref{eq3}) to be written as follows:
\begin{align}\label{eq4}
{\mathcal{L}}(f_{\mathbf{w}})=\sum\limits_{i = 1}^M {\frac{1}{2}{{\left\| {{\bf{e}}_i^{que}}(n) \right\|}^2}}
\end{align}
where ${\bf{e}}_i^{\rm{que}}(n) = {\bf{d}}_i^{\rm{que}}(n) - [({\bf{w}}_i'(n))^{\rm{T}} {\bf{X}}_i^{\sup}(n)] *s(n)$; ${\left\| \cdot \right\|}$ is the $l_2$ norm. The update of $\mathbf{w}$ using the gradient descent method is computed as
\begin{align}\label{eq5}
{\bf{w}}(n) &={\bf{w}}(n-1)-\beta \sum_{\mathcal{T}_{i}\sim p(\mathcal{T})} \nabla_{{\bf{w}}} \mathcal{L}_{\mathcal{T}_{i}}(f_{{\bf{w}}_i'(n)})\\
&={\bf{w}}(n-1) +\beta \sum\limits_{i = 1}^M \nabla_{{\bf{w}}} {\frac{1}{2}{{\left\| {{\bf{e}}_i^{que}}(n) \right\|}^2}} \nonumber
\end{align}
where $\beta$ is the step size for the outer-layer model. With the generic approaching method in the ANC area \cite{FxLMS,ANC1 ,FxNLMS,RFCGFxGHT}, (\ref{eq5}) can be further written as
\begin{flalign}\label{eq6}
\hspace{-0.5cm} {\bf{w}}(n)={\bf{w}}(n-1) + \beta \sum\limits_{i = 1}^M  \left[{\bf{\tilde Q}}_i^{\sup}(n) ({\bf{\tilde X}}_i^{\rm{que}}(n) {({\bf{e}}_i^{\rm{que}}(n))^{\rm{T}}} ) \right] \hspace{-0.4cm}
\end{flalign}
where ${\bf{\tilde X}}_i^{\rm{que}}(n)= [{\bf{\tilde x}}_{i,1}^{\rm{que}}(n),\cdots,{\bf{\tilde x}}_{i,J}^{\rm{que}}(n)]$ with its $j$th column being ${\bf{\tilde x}}_{i,j}^{\rm{que}}(n)={\bf{x}}_{i,k}^{\rm{que}}(n)*\hat{s}(n)$; ${\bf{\tilde Q}}_i^{\sup}(n)={\bf{E}} - \alpha {\bf{\tilde R}}_i^{\sup}(n) $ with ${\bf{\tilde R}}_i^{\sup}(n) \in {{\mathbb{R}}}{^{L \times L}} $ being the sum of covariance matrices of $\{{\bf{x}}_{i,k}^{\sup}(n) \}_{k=1}^K$, i.e., ${\bf{\tilde R}}_i^{\sup}(n)  = \sum\limits_{k = 1}^K {{\bf{\tilde x}}_{i,k}^{\sup }(n)} {({\bf{\tilde x}}_{i,k}^{\sup }(n))^{\rm{T}}}$. Combining (\ref{eq2})$-$(\ref{eq6}), the MAML-FxLMS algorithm is derived and summarized in Algorithm \ref{MAML-FxLMS}.
\begin{algorithm}[!t]\small
\caption{The MAML-FxLMS algorithm.} \label{MAML-FxLMS}
{\begin{algorithmic}
\STATE \hspace{-0.5cm}\textbf{Initialization}: $p(\mathcal{T})$: distribution over tasks;
$\alpha, \beta>0$: step size
\\\hspace{-0.5cm}hyperparameters; ${\bf{w}}(0)={\bf{0}}$: initial weight
\\\hspace{-0.5cm}\textbf{While} $n=1,2,\cdots$ \textbf{do}
\\\ \ \ \hspace{-0.5cm}\textbf{for all tasks $\mathcal{T}_{i} \sim p(\mathcal{T})$} \textbf{do}
\\\ Transmit $\mathbf{w}(n-1)$ into the inner-layer, i.e., ${\bf{w}}_i(n)={\bf{w}}(n-1)$
\\\ Sample $K$ input-output pairs $\left\{ {{\bf{x}}_{i,j}^{\sup}(n),d_{i,j}^{\sup}(n)} \right\}_{j = 1}^K$ from $D_i^{\sup}$
\\\ Compute the error vector by
\\\ \ \ \hspace{1cm}${\bf{e}}_i^{\sup}(n) = {\bf{d}}_i^{\sup}(n) - [{{\bf{w}}_i^{\rm{T}}}(n){\bf{X}}_i^{\sup}(n)] * s(n)$
\\\ Update the filter for the inner-layer model by (\ref{eq2})
\\\ Sample $J$ input-output pairs $\left\{ {{\bf{x}}_{i,j}^{\rm{que}}(n),d_{i,j}^{\rm{que}}(n)} \right\}_{j = 1}^J$ from $D_i^{\rm{que}}$
\\\ \ \ \hspace{-0.5cm}\textbf{end for}
\\\hspace{-0.2cm}Update the outer-layer model by (\ref{eq6})
\\\hspace{-0.5cm}\textbf{end}
\end{algorithmic}}
\end{algorithm}
Unlike the MAML method in \cite{meta1}, which updates the model over a randomly grabbed batch of tasks at each iteration, the MAML-FxLMS algorithm proposed in this paper updates the model over all tasks, but updating model with a batch of tasks is a simple extension and is usually adopted when the number of tasks is large.

The MAML-FxLMS algorithm aims to enable the pre-trained control filter to adapt to the new unseen homologous noise, i.e., a new task, and thus, the desired control filter is not an optimal control filter for a task but a control filter that can rapidly convergence on new tasks with the gradient-based learning rule. Finally, we can obtain $N$ control filters trained by MAML-FxLMS for all $N$ categories, which constitutes the control filter database.

\section{Filter selection and noise cancellation}
By using the MAML-FxLMS algorithm, the control filter database is established in the last section. The correct category of the arriving reference noise should be justified before cancellation. On the one hand, a one-dimensional audio clip shows only its amplitude change in time, while its spectrogram, such as the Mel or MFCC spectrograms, can retain both time and frequency domain information. That allows the classification model to capture both the dynamics of the audio over time and frequency, improving the classification accuracy. On the other hand, ResNet has been widely used in image classification and proven to have good generalization performance over recognition tasks \cite{ResNet}.
Therefore, a 50-layer ResNet version 2 (ResNet-50v2) is introduced in this paper to classify the noise category. When the ANC system is activated, a short noise segment of the reference noise is picked up to plot its Mel spectrogram, which is immediately input into ResNet-50v2.

Then, the best-matched control filter is chosen according to the classification result and adopted as the initial filter in the next module.
In the noise cancellation module, we update the filter with the basic FxLMS algorithm, which can test the fast learning and tracking abilities of the pre-trained control filters for new tasks intuitively.

\section{Simulations}
Simulations based on a popular common noise dataset, i.e., ESC-50 \cite{ESC-50}, are performed in this section. The classification performance of the classifiers within the proposed and other comparison models over the ESC-50 dataset is first verified and compared. Then, the control filter databases built by different models using different methods are established. Finally, simulation experiments for single category and cross-category noise cancellation are carried out to test the superiorities of the proposed MAML-FxLMS algorithm in terms of convergence speed, tracking and steady-state noise cancellation abilities. Averaged noise reduction (ANR) \cite{Survey,RFCGFxGHT} calculated by
${\rm{ANR}}(n) = 20{{\rm{log}}_{10}}({{A_{e}(n)}}/{{{A_{d}(n)}}})$ is adopted as evaluation index in noise cancellation module, with ${A_{e}(n)}$ and ${A_{d}(n)}$ denoting the estimates of absolute values of the residual and the disturbance noise, respectively.
\subsection{Noise classification over ESC-50 dataset}
ESC-50 dataset \cite{ESC-50} constituted by $2000$ labeled environmental noise recordings gathered from $50$ different subclasses is divided into $5$ major categories, including \emph{Animal}, \emph{Natural}, \emph{Human}, \emph{Interior}, and \emph{Exterior}. Each recording lasts $5s$ and has a sampling rate of $44.1$kHz. The less noise data required by the classifier, the better for the online application of the model, so each recording is trimmed into two $2.5s$ recordings for training the classifier.
To fully exploit the feature extraction and representation capabilities of deep neural network classifiers, the data augmentation methods are adopted by incorporating the following transformations:
\begin{itemize}
  \item Time scaling: The duration of each audio was stretched with factor 1.5 or compressed with factor 0.8.
\end{itemize}
\begin{itemize}
  \item Pitch shifting: The tone of each audio was pitched up with factor 4.5 or pitched down with factor -4.5 in semitones.
\end{itemize}

By incorporating these two transformations, more $16000$ noise recordings based on four different data augmentation methods with $[1.5, 4.5]$, $[1.5, -4.5]$, $[0.8, 4.5]$, and $[0.8, -4.5]$ are generated. A total of $20000$ recordings are used for classification, wherein $1200$ raw recordings constitute the test-set. The 2-dimensional convolutional neural network (2-D CNN) with two convolutional layers, a batch normalization layer, and two fully-connected layers with respectively $512$ and $50$ neurons, typically used in the SFANC algorithm \cite{SelectiveCNN} is compared. The ResNet-50v2 and 2D CNN \cite{SelectiveCNN} classifiers are both trained with ``categorical crossentropy'' loss, ``RMSprop'' optimizer, and a learning rate of $5 \times {10^{ - 5}}$.

\begin{figure}
	\subfigure[]{
        \hspace{-0.35cm}
		\label{fig31}
		\begin{minipage}[b]{0.24\textwidth}
			\centering
			\includegraphics[width=45mm,height=35mm]{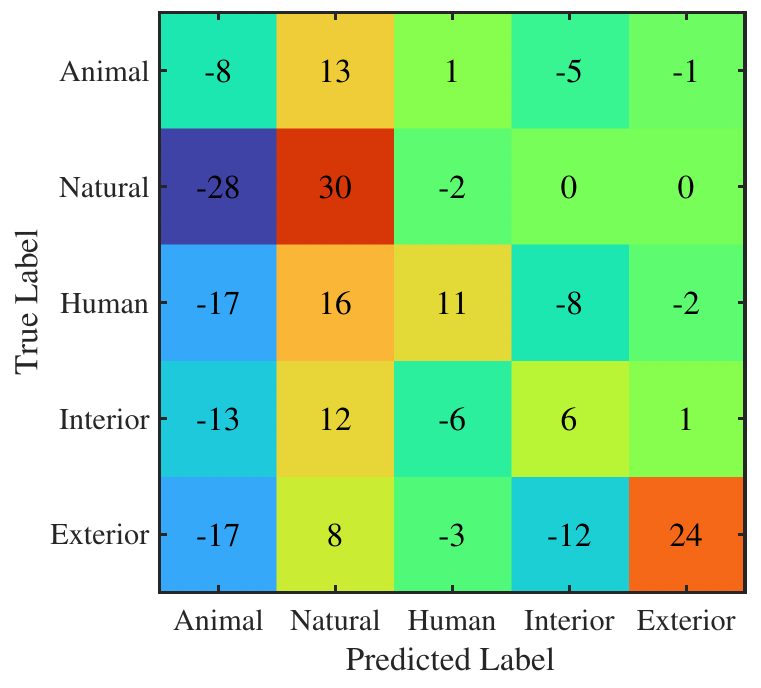}
	\end{minipage}}%
	\subfigure[]{
		\label{fig32} %
			\centering
			\includegraphics[width=46mm,height=35mm]{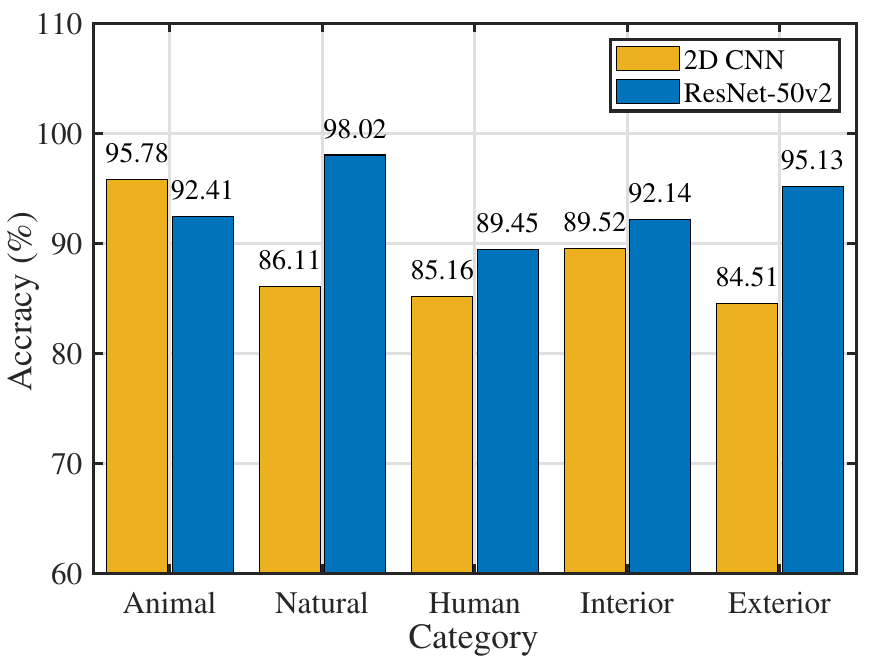}
}
\caption{ (a) Difference between the confusion matrices obtained by ResNet-50v2 and 2D CNN classifiers: positive/negative values along the diagonal mean the classification accuracy is increased/reduced, negative/positive values off the diagonal mean the confusion is reduced/increased. (b) The classification accuracy of ResNet-50v2 and 2D CNN over $5$ major categories.}
\label{Figure44}
\vspace{-5pt}
\end{figure}

Fig. \ref{Figure44} and Table \ref{tab1} show the classification results over ESC-50. The difference between the confusion matrices yielded by ResNet-50v2 and 2D CNN classifiers for the $5$-category classification task and their exact classification accuracy are shown in Figs. \ref{fig31} and \ref{fig32}, respectively. The ``Accuracy 1'' and ``Accuracy 2'' mentioned in Table \ref{tab1} are computed by averaging accuracy over the $50$ subclasses and the $5$ major categories, respectively. From Table \ref{tab1} and Fig. \ref{fig32}, we see that the classification performance of ResNet-50v2 is superior to that of 2D CNN, with improvements by $5.75\%$ and $5.25\%$ in ``Accuracy 1'' and ``Accuracy 2'', respectively.

\begin{table}[!t]
\caption{Model parameters and classification accuracy of ResNet-50v2 and 2D CNN classifiers.}\label{tab1}
\centering
\begin{threeparttable}
\setlength{\tabcolsep}{1mm}{
\begin{tabular}{c c c c c c}
\toprule
{ESC-50}\\
\hline
{Classifier} &{Epochs} &{Batch size} &{Augmentation} &{Accuracy 1}  &{Accuracy 2}\\
\midrule
{2D CNN}&{$20$} &{$16$} &{True} &{$84.75\%$} &{$88.17\%$} \\
{ResNet-50v2}&{$20$} &{$16$} &{True} &{$90.50\%$} &{$93.42\%$} \\%
\bottomrule
\end{tabular}}
\end{threeparttable}
\vspace{-10pt}
\end{table}
\subsection{Dynamic time-varying noise cancellation}\label{Performance Verification}
Taking \emph{Animal} category, including \emph{dog}, \emph{rooster}, \emph{pig}, \emph{cow}, \emph{frog}, \emph{cat}, \emph{hen}, \emph{insects}, \emph{sheep}, and \emph{crow} subclasses, as an example, the noise data of its first nine subclasses is used for fixed-filter training by MAML-FxLMS algorithm, while the noise data of its tenth subclass is used for noise cancellation performance verification. The FxLMS, SFANC, and SFANC-FxNLMS algorithms are chosen as the comparison algorithms, and the fixed control filters of SFANC and SFANC-FxNLMS are the same and fully pre-trained using the noise data from the \emph{crow} subclass. The arriving reference noise is a $25s$ audio clip extracted from the \emph{crow} subclass. And the corresponding pre-trained control filter will be chosen as the initial filter when the noise is classified into a certain category by the classifier. In the following two cases, $\hat{S}(z)=S(z)=z^0+z^{-1}+z^{-2}+0.5z^{-3}$ and parameters $K=J=10$ are configured.
\begin{figure}
\vspace{-5pt}
	\subfigure[]{
		\label{fig1}
		\begin{minipage}[b]{0.24\textwidth}
			\centering
			\includegraphics[width=46mm,height=37mm]{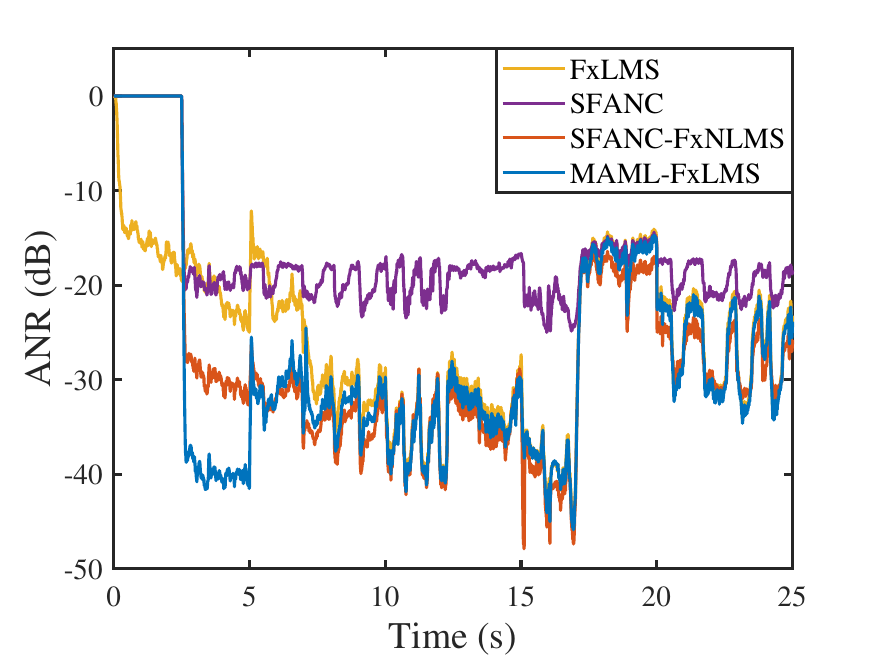}
	\end{minipage}}%
	\subfigure[]{
		\label{fig2} %
			\centering
			\includegraphics[width=46mm,height=37mm]{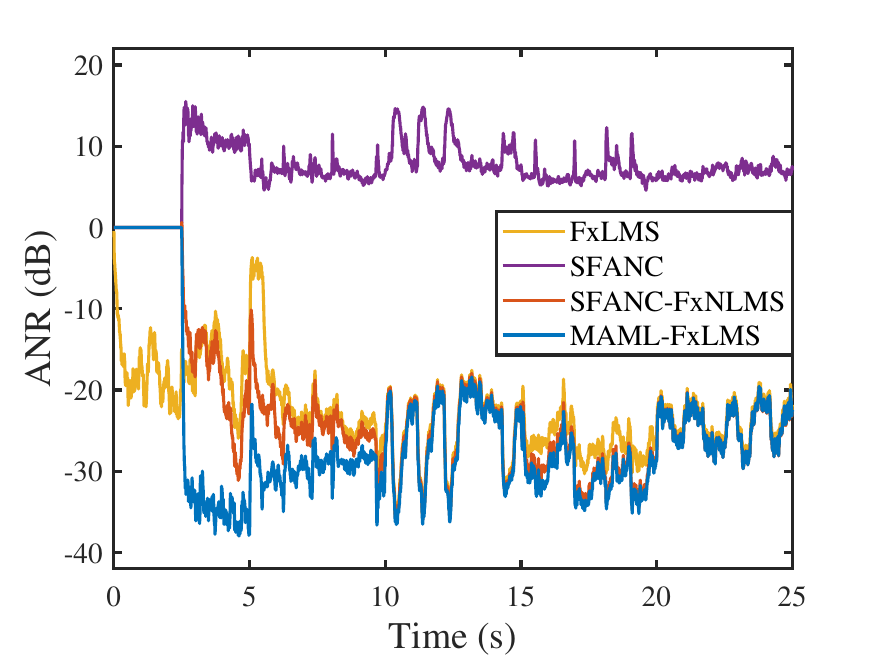}
}
\caption{ANR curves provided by algorithms for \emph{Animal} noise cancellation in: (a) case 1, (b) case 2.}
\label{Figure4}
\vspace{-8pt}
\end{figure}
\subsubsection{Case 1}\label{case1}
A short primary path $P(z)=1.5z^0+1.3z^{-1}-0.6z^{-2}-1.2z^{-3}-1.3z^{-4}+1.2z^{-5}$ with a length of $6$ is considered in this case. The filter length of all algorithms is set as $L=10$. In the pre-training module, the step sizes $\alpha=\beta=0.03$ are used for MAML-FxLMS, and $\mu=0.001$ is used for SFANC and SFANC-FxNLMS. In noise cancellation module, to make all algorithms' convergence as fast as possible, the step size $\mu=0.02$ is set for FxLMS, SFANC-FxNLMS, and MAML-FxLMS algorithms.
\subsubsection{Case 2}\label{case2}
A long primary path with a length of $64$ generated by a Gaussian distribution, i.e., ${\cal N}(0,0.1)$, is used in this case. The filter length of all algorithms is set as $L=72$. In pre-training module, the step sizes $\alpha=\beta=0.0012$ are set for MAML-FxLMS, and $\mu=0.03$ is set for SFANC and SFANC-FxNLMS. In noise cancellation module, the step sizes are set as 0.001, 0.0015, and 0.06 for FxLMS, SFANC-FxNLMS, and MAML-FxLMS algorithms, respectively, for fair comparison.

Suppose the arriving reference noise is classified into the correct category. The obtained ANR curves in \emph{Case 1} and \emph{Case 2} are plotted in Figs. \ref{fig1} and \ref{fig2}, respectively. The SFANC based algorithms, including SFANC, SFANC-FxNLMS, and MAML-FxLMS, require a $2.5s$ online data acquisition for noise classification. Form Fig. \ref{Figure4}, we see that (\rmnum{1}) although FxLMS does not require data acquisition, it needs a long convergence process; (\rmnum{2}) SFANC cannot achieve desired noise cancellation or even divergence when faced with a new unseen audio distinct from the pre-trained audio; (\rmnum{3}) although SFANC-FxNLMS keeps updating continuously, it still needs a relatively long time to adapt to a new noise cancellation task, because it merely utilizes a random audio segment to pre-train the initial filter and cannot learn the statistical characteristics of other audios belong to the same subclass; (\rmnum{4}) MAML-FxLMS leverages the two-layer structure of meta-learning, enabling it to learn the statistical characteristics of noises from the same category and achieve a steady state within just a few iterations after classification.
\subsection{Non-stationary noise cancellation}
Consider the human-movement in practice, a non-stationary noise environment that includes a transition from \emph{Exterior} to \emph{Human} categories at $15s$ is constructed to verify the tracking ability of the proposed MAML-FxLMS algorithm. Similar to the pre-training process introduced in Section \ref{Performance Verification}, the initial filters for \emph{Exterior} and \emph{Human} noises are learned with $\alpha=\beta=0.002$ and $\alpha=\beta=0.005$ for MAML-FxLMS, and with step sizes $0.02$ and $0.01$ for SFANC and SFANC-FxNLMS. The lengths of the primary path and the control filter are set as the same as that in Section \ref{case1}. The step size $\mu=0.015$ is set for FxLMS, SFANC-FxNLMS, and MAML-FxLMS in noise cancellation module.

\begin{figure}
\vspace{-5pt}
	\subfigure[]{
		\label{fig3}
		\begin{minipage}[b]{0.24\textwidth}
			\centering
			\includegraphics[width=46mm,height=36mm]{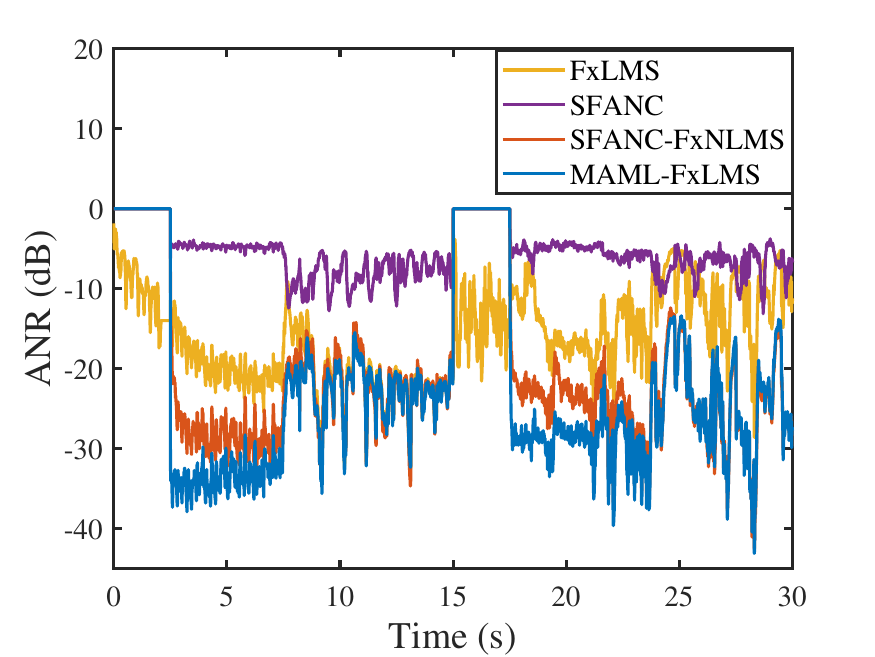}
	\end{minipage}}%
	\subfigure[]{
		\label{fig4} %
			\centering
			\includegraphics[width=46mm,height=36mm]{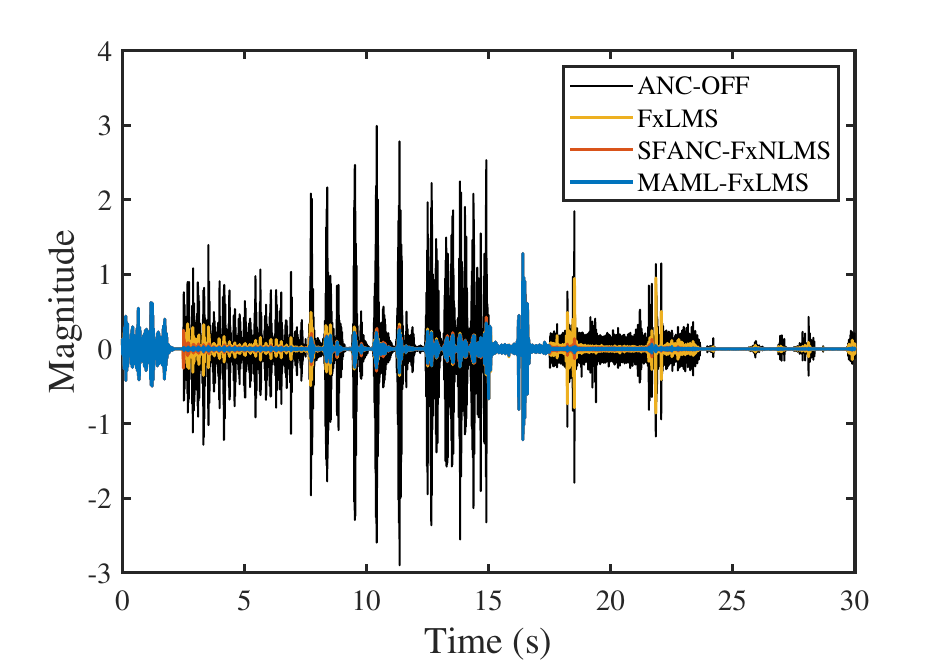}
}
\caption{(a) ANR curves obtained in non-stationary noise environment, (b) the corresponding residual error signals.}
\label{Figure5}
\vspace{-8pt}
\end{figure}

Figs. \ref{fig3} and \ref{fig4} plot the ANR curves and the corresponding residual errors of all comparison algorithms, respectively. It is obviously that the FxLMS tends to fall into suboptimal solutions when dealing with strong time-varying noise. Since SFANC rarely learns the information of a random audio clip extracted from the same subclass to obtain the control filter, its noise cancellation ability is quite limited. Moreover, even though SFANC-FxNLMS provides the adaptive learning ability to track the dynamic time-varying noise, it requires a long learning time, which is unacceptable in the real-world. The fast tracking ability of MAML-FxLMS is surprising and meaningful for real-world noise cancellation. As Fig. \ref{fig4} shows, the proposed MAML-FxLMS can also obtain the minimum residual errors for both \emph{Exterior} and \emph{Human} noises.
\raggedbottom
\section{Conclusion}
In this paper, the meta-learning is introduced into the selective fixed-filter active noise control (SFANC) system to pre-train the fixed control filters for the first time, generating the modified model-agnostic meta-learning filtered-x least mean square (MAML-FxLMS) algorithm. One of the biggest highlights is that the MAML-FxLMS learns not the optimal control filer for a noise cancellation task,  but the control filter can quickly adapt to the new noise cancellation task. Simulations on the ESC-50 noise dataset show that the proposed SFANC system is more feasible and achieves superior noise attenuation performance in terms of tracking and steady-state noise cancellation abilities.
Note that all the SFANC based methods require the online acquisition of a noise frame to categorize the arriving reference noise, and the length of the noise frame required by noise classifiers determines the initial response speed of these SFANC systems. Therefore, further improving the classification performance of the noise classifier with a much shorter noise frame in the future is noteworthy.

\balance

\end{document}